\documentstyle[twoside,fleqn,espcrc2,epsf]{article}

\title{Spectral Functions of Hadrons in Lattice QCD\thanks{Talk
 given by Y. Nakahara and M. Asakawa at LATTICE99.}}

\author{Y. Nakahara$^a$, M. Asakawa\address{Department of Physics,Nagoya
University, Nagoya 464 - 8602, Japan}, and T. Hatsuda\address{Physics
Department, Kyoto University, Kyoto 606-8502, Japan}}

\begin{document}
\begin{abstract}
 Using the maximum entropy method, spectral functions of
  the pseudo-scalar
 and vector mesons are extracted from lattice Monte Carlo
 data of the imaginary time Green's functions.
 The resonance and continuum structures as well as the
 ground state peaks are successfully obtained. 
 Error analysis of the resultant spectral functions is
 also given on the basis of  the Bayes probability theory.

\end{abstract}

\maketitle

\section{Introduction}
 
 The spectral functions (SPFs) of hadrons play a special role in
 physical observables in QCD (See the examples
 in  \cite{shuryak,negele}).
 However, the
 lattice QCD simulations so far 
 have difficulties in accessing the dynamical quantities 
 in the Minkowski space, because
 measurements on the lattice can only be carried out
 for discrete points in imaginary time.
 The analytic continuation from the imaginary time to the
 real time using the noisy lattice data 
 is highly non-trivial  and is even classified as an ill-posed
 problem. 
 
  Recently we made a first serious  attempt to
 extract SPFs of hadrons from lattice
 QCD data without making a priori assumptions on the spectral shape
 \cite{nah}.
  We use the 
 maximum entropy method (MEM), which has been successfully applied for 
 similar problems in quantum Monte Carlo simulations in
 condensed matter physics,
 image reconstruction in crystallography and 
 astrophysics, and so forth  \cite{physrep,linden}.
 In this report, we present  the
 results for the pseudo-scalar (PS) and
 vector (V) channels at $T=0$ using the continuum kernel
 and the lattice kernel of the integral transform.
 The latter analysis has not been reported in \cite{nah}.

\section{Basic idea of MEM}

The Euclidean correlation function $D(\tau)$ of an 
operator ${\cal O}(\tau,\vec{x})$  and its spectral decomposition
at zero three-momentum read
\begin{eqnarray}
D(\tau ) &=& \int 
\langle {\cal O}^{\dagger}(\tau,\vec{x}){\cal O}(0,\vec{0})\rangle d^3 x\label{KA}\\\nonumber
       &=&  \int_{0}^{\infty} \!\! K(\tau, \omega) A(\omega ) d\omega,
\end{eqnarray}
where $\tau > 0$, $\omega$ is a real frequency, and
$A(\omega)$ is SPF
(or sometimes called the {\em image}),
which is positive semi-definite.
The kernel $K(\tau, \omega)$  is proportional to
the Fourier transform of a free boson propagator with
mass $\omega$: At $T=0$ in the continuum limit, 
 $K = K_{cont}(\tau, \omega)
=\exp(-\tau\omega)$. 

Monte Carlo simulation provides  $D(\tau_i)$ 
on  the discrete set of temporal points $0 \le \tau_i /a \le N_\tau$.
From this  data with statistical noise, we need to reconstruct the 
spectral function $A(\omega)$ with continuous variable $\omega$. 
 This is a
typical ill-posed problem, where the number of data is much smaller than
the number of degrees of freedom to be reconstructed.
This makes the standard 
likelihood analysis and its variants
inapplicable \cite{others} unless strong assumptions
on the spectral shape are made.
MEM is a method to circumvent this difficulty
 through Bayesian  statistical inference of the most probable {\em image}
 together with its reliability \cite{physrep}.

  MEM
 is based on the Bayes' theorem in probability theory:
 $P[X|Y] = P[Y|X]P[X]/P[Y]$,
 where $P[X|Y]$ is the conditional probability of $X$ given $Y$.
 The most probable image 
 $A(\omega )$ for given lattice data $D$ is obtained by 
 maximizing the conditional probability 
 $P[A|DH]$, where $H$ summarizes all the
 definitions and
 prior knowledge such as  $A(\omega) \ge 0$.
 By the Bayes' theorem, 
\begin{equation}\label{bayes_latt}
P[A|DH] \propto P[D|AH]P[A|H] ,
\end{equation}
where $P[D|AH]$ ($P[A|H]$) is called the likelihood function
 (the prior probability).

For the likelihood function, 
 the standard $\chi^2$ is adopted, namely 
$P[D|AH]= Z_L^{-1} \exp (-L)$ with 
\begin{eqnarray}
&&\mbox{\hspace{-0.5cm}}L  = {1 \over 2} \sum_{i,j}
(D(\tau_i)-D^A(\tau_i)) \\[-0.3cm]
&&\mbox{\hspace{2.5cm}} \times \, C^{-1}_{ij}
(D(\tau_j)-D^A(\tau_j)).\nonumber \label{chi2}
\end{eqnarray}
$Z_L$ is a normalization factor given by
$Z_L = (2\pi)^{N/2} \sqrt{\det C}$ with
$N={\tau}_{max}/a - {\tau}_{min}/a+1$.
$D(\tau_i )$ is the lattice data averaged over gauge configurations 
and $D^A(\tau_i )$ is the correlation function defined 
by the right hand side of (\ref{KA}).
$C$ is an $N \times N$ covariance matrix of the data
 with $N$ being the number of temporal points to be used
 in the MEM analysis.
The lattice data have generally strong correlations among
different $\tau$'s, and it is essential to take into account the
off-diagonal components of $C$.

 Axiomatic construction  as well as intuitive "monkey
argument" \cite{skilling} show that, for positive distributions
such as SPF, the prior probability can be written
with parameters $\alpha$ and $m$ as
$P[A|H\alpha m]= Z_S^{-1} \exp (\alpha S)$. Here
$S$ is the Shannon-Jaynes entropy,
\begin{eqnarray}
&&\mbox{\hspace{-.5cm}}S =\\
&&\mbox{\hspace{-.5cm}} \ \ \int_0^{\infty} \left [ A(\omega ) - m(\omega )
- A(\omega)\log \left ( \frac{A(\omega)}{m(\omega )} \right ) \right ]
d\omega .\nonumber
\end{eqnarray}
$Z_S$ is a normalization factor:
$Z_S \equiv  \int e^{\alpha S} {\cal D}A$.
$\alpha$ is a real and positive parameter and 
$m(\omega )$ is a real function called the default model.
 
In the state-of-art  MEM \cite{physrep},  
the output image $A_{out}$ is
given by a weighted average over $A$ and  $\alpha$:
\begin{eqnarray}
&&\mbox{\hspace{-.5cm}}A_{out}(\omega) \nonumber \\
&&= \int A(\omega)  \ 
P[A|DH\alpha m]P[\alpha|DHm] \  {\cal D} A  \ d\alpha \nonumber \\
&& \simeq  \int A_{\alpha}(\omega) \   P[\alpha|DHm]  \ d\alpha .
\label{final}
\end{eqnarray}
Here $A_{\alpha}(\omega)$ is obtained by
maximizing the "free-energy"
\begin{eqnarray}
Q \equiv \alpha S - L,
\end{eqnarray}
 for a given $\alpha$. Here we assumed 
that $P[A|DH\alpha m] $ is
sharply peaked around $A_{\alpha}(\omega)$.
$\alpha$ dictates the relative weight of the
entropy  $S$ (which tends to fit $A$ to the default model $m$)
and the likelihood function $L$ (which tends to fit $A$ to the
lattice data). Note, however, that 
$\alpha$ appears only in the intermediate
step and is  integrated out in the final result.
 Our lattice data show that the weight factor $P[\alpha|DHm]$, which  
is calculable using $Q$ \cite{physrep},
is highly peaked around its maximum
$\alpha = \hat{\alpha}$.
 We have also studied the stability of the 
$A_{out}(\omega)$ 
against a reasonable variation of $m(\omega)$.

The non-trivial part of the MEM analysis is to find a global
maximum of $Q$ in the functional space of $A(\omega)$,
which has typically 750 degrees of freedom in our case. 
We have utilized the singular value decomposition (SVD) of
 the kernel  
to define the search direction in this functional space.
 The method works successfully to find the global maximum
within reasonable iteration steps.

\section{MEM with mock data}

To check our MEM code and to see the dependence of the 
MEM image on the quality 
of the data, we made the following test using
mock data.
(i) We start with an input  image 
$A_{in}(\omega) \equiv \omega^2 \rho_{in}(\omega)$
in the $\rho$-meson channel which simulates
the experimental $e^+e^-$ cross section.
Then we calculate $D_{in}(\tau)$ from  $A_{in}(\omega)$ using eq.(\ref{KA}). 
(ii) By taking $D_{in}(\tau_i)$ at $N$
discrete points 
and adding a Gaussian noise, we create a mock data
$D_{mock}(\tau_i)$.
The variance of the noise $\sigma (\tau_i)$
is given by $\sigma (\tau_i)= b \times D_{in}(\tau_i) \times \tau_i /a$
with a parameter $b$, which controls the noise level \cite{noise}.
(iii) We construct the output image 
$A_{out}(\omega)  \equiv \omega^2 \rho_{out}(\omega)$
using MEM with $D_{mock}(\tau_{min} \le \tau_i \le \tau_{max})$ 
and compare the result with $A_{in}(\omega)$.
In this test, we have assumed that $C$ is
diagonal for simplicity. 

In Fig.1, we show $\rho_{in}(\omega)$, and 
$\rho_{out} (\omega)$  for two
sets of parameters, (I) and  (II).
As for $m$, we choose a form 
$m(\omega) = m_0 \omega^2$ with $m_0 = 0.027$, which is
motivated by the
asymptotic behavior of $A$ in perturbative QCD,
$A(\omega \gg 1 {\rm GeV})
= (1/4 \pi^2) (1+\alpha_s / \pi) \omega^2$.
The final result is, however, insensitive to 
the variation of $m_0$ even by factor 
5 or 1/5. The calculation of $A_{out}(\omega)$
has been done by discretizing the $\omega$-space
with an equal  separation of 10 MeV between adjacent points.
This number is chosen for the reason we 
shall discuss below.
The comparison of the dashed line (set (I)) and
the dash-dotted line (set (II))
shows that increasing $\tau_{max}$ 
and reducing the noise level $b$ lead to better
SPFs closer to the input SPF. 

We have also checked that MEM can nicely reproduce
other forms of the mock SPFs.  In particular, 
MEM works very well
to reproduce not only the broad structure but also
the sharp peaks close to the delta-function as far as 
the noise level is sufficiently small.

\begin{figure}[h]
\epsfxsize=8.2cm
\centerline{\epsfbox{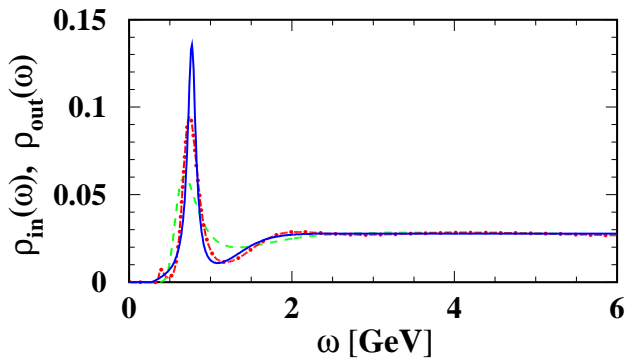}}
\vskip0mm
\caption{
 The solid line is $\rho_{in}(\omega)$.
 The dashed line and dash-dotted line are $\rho_{out}(\omega)$
 obtained with  parameter set (I) $a = 0.0847$ fm, 
 $1 \le \tau /a \le 12$, $b=0.001$ 
 and set (II) $a = 0.0847$ fm, 
 $1 \le \tau /a \le 36$, $b=0.0001$, respectively. }
\label{fig1}
\end{figure}

\section{MEM with lattice data}

 To apply MEM to actual lattice data, 
 quenched lattice QCD simulations
have been done 
with the plaquette gluon action and the Wilson quark action
by the open MILC code with minor modifications \cite{milc}.
The lattice size is $20^3 \times 24$
with $\beta =6.0$, which corresponds to 
$ a = 0.0847$ fm ($a^{-1} = 2.33$ GeV),
$\kappa_c  = 0.1571$ \cite{kc}, and the
spatial size of the lattice $L_s a = 1.69 $ fm.
Gauge configurations are generated by the heat-bath and
over-relaxation algorithms with a ratio $1:4$. Each configuration
is separated by 1000 sweeps.
Hopping parameters are chosen to be
$\kappa =$ 0.153, 0.1545, and 0.1557 with $N_{conf}=161$
for each $\kappa$.
For the quark propagator, the Dirichlet (periodic)
boundary condition
is employed for the temporal (spatial) direction.
 We have also done the simulation with periodic
 boundary condition in the temporal direction and
 obtained qualitatively the same results.
To calculate the two-point correlation functions,
we adopt a point-source at $\vec{x}=0$  and a point-sink
averaged over the spatial lattice-points.

We use data at $1 \le \tau_i/a  \le 12 (24) $ 
for the Dirichlet (periodic) boundary condition in the 
 temporal direction.
To avoid the known pathological behavior of the eigenvalues of 
$C$ \cite{physrep}, we take $N_{conf} \gg N$.

We define SPFs for the PS and V channels as
\begin{equation}
 A(\omega) = \omega^2 \rho_{_{PS},_{V}}(\omega) ,
\end{equation}
so that $\rho_{_{PS,V}}(\omega \rightarrow {\rm large}) $ approaches
a finite constant as predicted by  perturbative QCD.
For the MEM analysis,
we need to discretize the $\omega$-integration in (\ref{KA}).
Since $\Delta \omega$ (the mesh size) $\ll 1/\tau_{max}$
should be satisfied  to suppress the discretization error,
we take $\Delta \omega$ = 10 MeV.
$\omega_{max}$ (the upper limit for the $\omega$ integration)  
should be comparable to the maximum
available momentum on the lattice:
$\omega_{max} \sim \pi /a \sim 7.3$ GeV.
We have checked that larger
values of $\omega_{max} $ do not change 
the result of $A(\omega)$ substantially, while smaller
values of $\omega_{max} $ distort the high energy end of the
spectrum. The dimension
of the image to be reconstructed  is $N_{\omega} \equiv
\omega_{max}/\Delta \omega \sim 750$,
which is in fact much larger than the maximum number of 
Monte Carlo data  $N = 25$.

In Fig.2 (a) and (b), we show the reconstructed images for
each $\kappa$ in the case of the Dirichlet boundary condition. 
 Here we use the continuum kernel $K_{cont} = \exp(-\tau \omega)$
 in the Laplace transform.
 In these figures, we have used $m = m_0 \omega^2$ with
$m_0 = 2.0 (0.86)$ for PS (V) channel motivated by
the perturbative estimate of $m_0$ 
(see eq.(\ref{cont-V}) and the text below).
We have checked that 
the result is not sensitive, within the statistical
significance of the image, to the variation
of $m_0$ by factor 5 or 1/5. 
The obtained images have a common structure:
the low-energy peaks corresponding to
$\pi$ and $\rho$, and the broad structure in the high-energy
region.  From the position of the pion peaks in Fig.2(a), we extract 
$\kappa_c = 0.1570(3)$, which is
consistent with $ 0.1571 $ \cite{kc} determined from the
asymptotic behavior of $D(\tau)$. 
The mass of the $\rho$-meson in the chiral limit
extracted from the peaks in Fig.2(b) reads
$m_{\rho}a = 0.348(15)$. This  is also consistent with 
$m_{\rho}a = 0.331(22) $  \cite{kc}
determined by the asymptotic behavior.
Although our maximum value of the
fitting range $\tau_{max}/a =12$ 
marginally covers the asymptotic limit in $\tau$, we 
can extract reasonable masses for $\pi$ and $\rho$.
The width of $\pi$ and $\rho$  in Fig.2
is an artifact due to the statistical errors of the
lattice data. In fact, in the quenched approximation, there is no room 
for the $\rho$-meson to decay into two pions.

As for the second peaks in the PS and V channels,
the error analysis discussed in Fig.4 shows that
their spectral ``shape" does not have much  statistical
significance, although the existence of the
non-vanishing spectral strength is significant.
Under this reservation, we fit the position of the 
second peaks and made linear
extrapolation to the chiral limit with 
the results, $m^{2nd}/m_{\rho} = 1.88(8)
(2.44(11))$  for the PS (V) channel.
These numbers should be compared with the 
experimental values:
$m_{\pi(1300)}/m_{\rho} = 1.68$,
and $m_{\rho(1450)}/m_{\rho} = 1.90$
or $m_{\rho(1700)}/m_{\rho} = 2.20$.

One should remark here that, 
in the standard two-mass fit of $D(\tau)$, the mass
of the second resonance is highly sensitive to the 
lower limit of the fitting range, e.g., 
$m^{2nd}/m_{\rho} = 2.21(27) (1.58(26))$  for $\tau_{min}/a = 8 (9)$
in the $V$ channel with $\beta=6.0$ \cite{kc}.
This is because the contamination from the short distance
contributions from $\tau < \tau_{min}$ is not
under control in such an approach.
On the other hand, MEM does not suffer from this difficulty 
and can utilize the full
information down to $\tau_{min}/a=1$.
Therefore, MEM opens
a possibility of systematic study of higher resonances 
with lattice QCD data.

\begin{figure}[h]
\vskip4mm
\epsfxsize=8.2cm
\centerline{\epsfbox{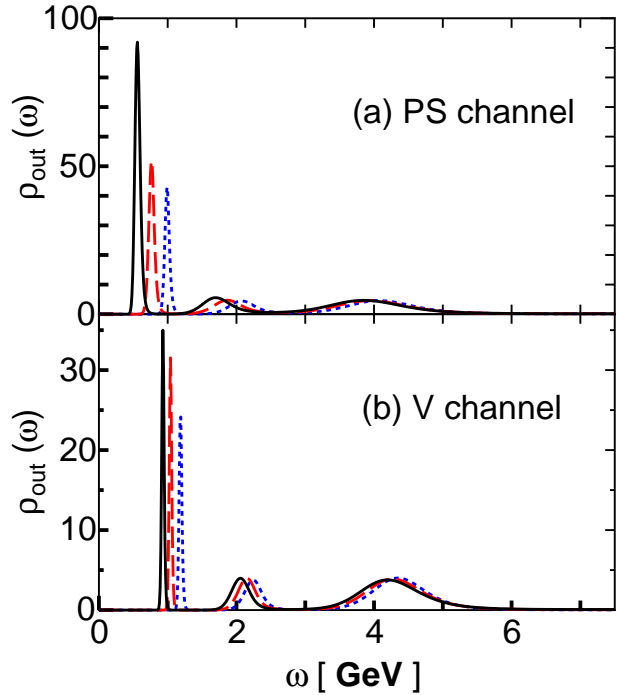}}
\vspace{-0.5cm}
\caption{
	Reconstructed image $\rho_{out}(\omega)$ for
the PS (a) and V (b) channels. The solid, dashed, and dotted
lines are for $\kappa=$ 0.1557, 0.1545, and 0.153, respectively.
For the PS (V) channel, $m_0$ is taken
to be 2.0 (0.86). $\omega_{max}$ is 7.5 GeV 
in this figure and Fig.3.}
\label{fig2}
\end{figure}

As for the third bumps in Fig.2, the spectral ``shape" 
is statistically not significant as is discussed in Fig.4,
and they should rather be considered a part of the
perturbative continuum instead of a single resonance.
Fig.2 also shows that SPF decreases substantially
above 6 GeV; MEM automatically detects
the existence of the momentum cutoff on the lattice $\sim \pi/a$.
It is expected that MEM with the data on finer lattices leads to larger
ultraviolet cut-offs in the spectra.
The  height of the asymptotic form of the
spectrum at high energy is estimated as 
\begin{eqnarray}
&&\mbox{\hspace{-.5cm}} \rho_{_V}(\omega \simeq 6 {\rm GeV})\\
&&\mbox{\hspace{.5cm}} = {1 \over 4 \pi^2} \left ( 1 + {\alpha_s \over \pi} \right )
 \left ( {1 \over 2\kappa  Z_{_V}} \right )^2 \nonumber
\simeq 0.86 . \nonumber
\label{cont-V}
\end{eqnarray}
The first two factors are the $q \bar{q}$
continuum expected from  perturbative QCD.
The third factor contains  the non-perturbative
renormalization constant for the lattice composite operator.
We adopt 
$Z_{_V} = 0.57$ determined from the two-point functions at
$\beta$ = 6.0 \cite{mm86} together with $\alpha_s = 0.21$
and $\kappa = 0.1557$.
Our estimate in eq.(\ref{cont-V}) 
is  consistent with the high energy part of the spectrum
in Fig.2(b) after averaging over $\omega$.
We made a similar estimate for the PS channel using
$Z_{_{PS}} = 0.49 $ \cite{shi} and obtained
$\rho_{_{PS}}(\omega \simeq 6 {\rm GeV}) \simeq 2.0$. This is
also consistent with Fig. 2(a).
We note here that 
an independent analysis of the imaginary time
correlation functions \cite{negele}
also shows that 
the lattice data at short distance is dominated by 
the perturbative continuum.

In Fig.3(a) and (b), the results using the 
 lattice kernel $K_{lat}$ are shown.
 $K_{lat}$ is obtained from the free boson propagator
 on the lattice. It reduces to $K_{cont}$ when $a \rightarrow 0$.
 The other parameters and boundary conditions are the
 same with Fig.2(a,b).
  The difference of Fig.2 and Fig.3 can be interpreted as
 a systematic error due to the finiteness of the 
 lattice spacing $a$. 
\begin{figure}[h]
%\vskip4mm
\epsfxsize=8.2cm
\centerline{\epsfbox{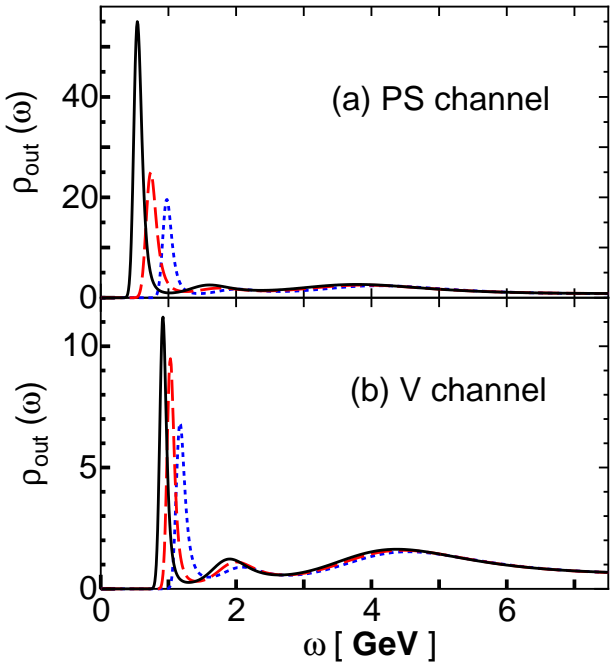}}
\vspace{-0.5cm}
\caption{
	Same with Fig.2 except for the use of the lattice
 kernel $K_{lat}$.}
\label{fig3}
\end{figure}

\section{Error analsis}

The statistical 
significance of the reconstructed image can be studied by the following
procedure \cite{physrep}.
Assuming that $P[A|DH\alpha m]$ has a Gaussian distribution
around the most probable image $\hat{A}$, we estimate the error 
by the covariance of the image, 
$- \langle (\delta_A \delta_A Q)^{-1} \rangle_{A=\hat{A}}$,
where $\delta_A$ is a functional derivative and
$\langle \cdot \rangle $ is an average over
a given energy interval.
The final error for $A_{out}$ is obtained 
by averaging the covariance over $\alpha$ with
a weight factor $P[\alpha |DHm]$.
Shown in Fig.4 is  the MEM image in the V channel for $\kappa= 0.1557$
with errors obtained in the above procedure. 
The height of each horizontal bar is
$\langle\rho_{out}(\omega)\rangle$
in each $\omega$ interval.
The vertical bar indicates the error of 
$\langle\rho_{out}(\omega)\rangle$.
The small error for the lowest peak in Fig.4
supports our identification of the peak with $\rho$. 
Although the existence of the non-vanishing
spectral strength of the 2nd peak and 3rd bump
is statistically significant, their spectral ``shape''
is either marginal or insignificant.
Lattice data with better quality are called for
to obtain better SPFs.

\begin{figure}[h]

\vskip4mm
\epsfxsize=8.2cm
\centerline{\epsfbox{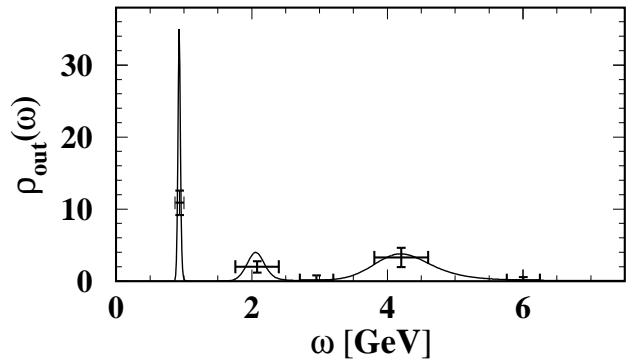}}
\vskip2mm
\caption{
	$\rho_{out}(\omega)$ in the $V$ channel 
 for $\kappa=$ 0.1557 with error attached.}
\label{fig4}

\end{figure}

\section{Summary}
We have made a first serious attempt to reconstruct
SPFs of hadrons from lattice QCD data.
We have used MEM, which allows us to study SPFs without
making a priori assumption on the spectral shape.
The method works well for the mock data and actual
 lattice data. 
 MEM produces resonance  and continuum-like structures
in addition to the ground state peaks.
The statistical significance of the image can be also analyzed. 
Better data with finer and larger lattice will
produce better images with smaller errors, and our
study is a first attempt towards this goal.

There are many problems which can be explored by MEM
 combined with lattice QCD data.  Some of the applications
 in the baryon excited states, hadrons at finite temperature, 
 and heavy quark systems will be reported in future
 publications \cite{nextpaper}.

We appreciate MILC collaboration for their open  codes
for lattice QCD simulations, which has enabled this research.
Our simulation was carried out on a Hitachi 
SR2201 parallel computer at Japan Atomic Energy Research Institute.
M. A. (T. H.) was partly supported by Grant-in-Aid for Scientific
Research No. 10740112 (No. 10874042) of the Japanese Ministry of Education,
Science, and Culture.

%%%%%%%%%%%%%%%%%%%%%%%%%%%%%%%%%%%%%%%%%%%%%%%%%%%%%%%%%%%%%%%%%%%

%%%%%%%%%%%%%%%%%%%%%%%%%%%%%%%%%%%%%%%%%%%%%%%%%%%%%%%%%%%%%%%%%%%


\begin{thebibliography}{99}

\bibitem{shuryak}
         E. V. Shuryak, Rev. Mod. Phys. {\bf 65},1 (1993).
\bibitem{negele}
        M. -C. Chu, J. M. Grandy, S. Huang, and J. W. Negele,
        Phys. Rev. D {\bf 48}, 3340 (1993).
\bibitem{nah}
       Y. Nakahara, M. Asakawa and T. Hatsuda,
       hep-lat/9905034 (Phys. Rev. D in press).
\bibitem{physrep} See the review,
         M. Jarrell and J. E. Gubernatis, Phys. Rep. {\bf 269},
	 133 (1996).
\bibitem{linden} 
        R. N. Silver et al., Phys. Rev. Lett. {\bf 65}, 496 (1990);
        W. von der Linden, R. Preuss, and W. Hanke,
        J. Phys. {\bf 8}, 3881 (1996);
        N. Wu, {\em The Maximum Entropy Method}, (Springer-Verlag,
        Berlin, 1997).

\bibitem{others}
        D. B. Leinweber, Phys. Rev. D {\bf 51}, 6369 (1995);
        D. Makovoz and G. A. Miller, Nucl. Phys. {\bf B468}, 293 (1996);
        C. Allton and S. Capitani, Nucl. Phys. {\bf B526}, 463 (1998);        
        Ph. de Forcrand et al., Nucl. Phys. B (Proc. Suppl.) {\bf 63A-C},
        460 (1998).
\bibitem{skilling} See e.g.,
        J. Skilling, 
        in {\em Maximum Entropy and Bayesian Methods},
        ed. J. Skilling (Kluwer, London, 1989), pp.45-52;
         S. F. Gull, ibid. pp.53-71.
\bibitem{noise}
        This formula is motivated by our lattice QCD data.  
\bibitem{milc} The MILC code ver. 5, \\ 
	http://cliodhna.cop.uop.edu/\~{}hetrick/milc .
\bibitem{kc}
        Y. Iwasaki et al., Phys. Rev. D {\bf 53}, 6443 (1996);
        T. Bhattacharya et al., ibid. 6486.
\bibitem{mm86}
	L. Maiani and G. Martinelli, Phys. Lett. {\bf B178}, 265 (1986).
\bibitem{shi}
        M. G\"{o}ckeler et al., Nucl. Phys. {\bf B544}, 699 (1999).
\bibitem{nextpaper} 
        M. Asakawa, T. Hatsuda and Y. Nakahara, in preparation.

\end{thebibliography}
\end{document}